\documentclass[hyper]{JHEP} 
\usepackage{epsfig}
\usepackage{amsmath}
 \usepackage{slashed}
\newcommand\fverb{\setbox\pippobox=\hbox\bgroup\verb}
\newcommand\fverbdo{\egroup\medskip\noindent%
            \fbox{\unhbox\pippobox}\ }

\newcommand\fverbit{\egroup\item[\fbox{\unhbox\pippobox}]}
\newbox\pippobox
\title{$(p,q)-$Five Brane and $(p,q)-$String  Solutions, Their
Bound State And Its Near Horizon Limit}

\author{
 Josef Kluso\v{n} \\
Institute for Theoretical Physics  and Astrophysics\\
Faculty of Science, Masaryk University\\
Kotl\'{a}\v{r}sk\'{a} 2, 611 37, Brno\\
Czech Republic\\
E-mail: \email{klu@physics.muni.cz}}

\abstract{We determine $(p,q)-$string and $(p,q)-$five brane
solutions of type IIB supergravity  using $SL(2,Z)-$symmetry  of the
full type IIB superstring theory. We also determine
$SL(2,Z)-$transformed solution corresponding to the bound state of
NS5-branes and fundamental strings. Then we analyze  its near
horizon limit and we show that it leads to the  $AdS_3\times S^3$
with mixed fluxes. } 
\def\det{\mathrm{det}}

\def\hx{\hat{x}}
\def\tr{\mathrm{Tr}}

\def\tx{\tilde{x}}

\def\mM{\mathcal{M}}

\def\hf{\hat{f}}

\def\tkappa{\tilde{\kappa}}

\def\bm{\mathbf{m}}
\def\bB{\mathbf{B}}

\newcommand{\hF}{\hat{F}}

\newcommand{\hH}{\hat{H}}

\newcommand{\mG}{\mathcal{G}}

\def\hr{\hat{r}}

\newcommand{\hg}{\hat{g}}

\newcommand{\hB}{\hat{B}}

\newcommand{\hJ}{\hat{J}}

\newcommand{\bH}{\mathbf{H}}

\begin{document}

\section{Introduction}
It is conjectured that the type IIB superstring theory possesses
$SL(2,Z)$  non-perturbative duality. The first evidence follows from
the manifestly $SL(2,R)-$invariance of type IIB supergravity
effective action \cite{Bergshoeff:1995as,Hull:1994ys} for recent
excellent review see \cite{Ortin:2015hya}. Further evidence follows
from the spectrum of non-perturbative objects that are presented in
type IIB theory:Dp-branes with $p=1,3,5,7,9$
\cite{Polchinski:1995mt}, fundamental string and NS5-brane
\footnote{For review, see
\cite{Polchinski:1998rr,Blumenhagen:2013fgp}.}. It was argued that
under S-duality fundamental string maps to D1-brane, D5-brane maps
to NS5-brane and so on. On the other hand we know that macroscopic
extend objects are sources of supergravity fields and hence it is
possible to find corresponding background solutions that solve the
supergravity equations of motions. Well known examples of such
solutions are fundamental string solution \cite{Dabholkar:1990yf} or
NS5-brane solution \cite{Callan:1991dj}. Then with the help of the
$SL(2,R)$-covariance of type IIB supergravity action new solutions
corresponding to $(p,q)-$string was found in \cite{Schwarz:1995dk}.
In fact, an existence of this solution serves as a further evidence
of $SL(2,Z)-$duality of type IIB string theory. The main idea of
this construction is to start with fundamental string solution and
performs $SL(2,R)$ rotation. Then the requirement that the resulting
configuration has to have integer charge in some units fixes entries
of this matrix as functions of these charges and asymptotic values
of dilaton and Ramond-Ramond one form. It is important to stress
that this solution depends on one harmonic function with manifestly
$SL(2,R)-$covariant coefficient.
 Then  this method
 was applied for the
construction of $(p,q)-$five brane backgrounds in \cite{Lu:1998vh}.
These very interesting backgrounds were analyzed recently from the
$(m,n)-$string probe point of view in \cite{Kluson:2016pxg} and it
was shown that when we perform $SL(2,R)$ rotation that maps the
macroscopic $(p,q)-$string background to the fundamental string
background  the probe string does not transform in the expected way
since now it carries non-integer charge with respect to NSNS two
form. The same situation also occurs in case of $(p,q)-$five brane
background. It was suggested there that the resolution of this
puzzle could be found when we consider a full type IIB superstring
theory that is invariant under $SL(2,Z)-$subgroup of $SL(2,R)$. This
paper is devoted to this analysis.

We propose that it is natural to search for the $(p,q)-$macroscopic
string solution with the presence of the source which is manifestly
$SL(2,R)$ covariant $(p,q)-$string action
\cite{Cederwall:1997ts,Tseytlin:1996it,Lozano:1997cy,Townsend:1997kr,Bergshoeff:2006gs}.
However the fact that there is no fractional string or D-brane
charges demands that the proper invariant group of type superstring
theory  is $SL(2,Z)$ rather than $SL(2,R)$ group and this group
\emph{should} be used for the construction of $(p,q)-$macroscopic
string and five brane solutions \footnote{Even if five brane
solutions are source free we can consider solutions where the source
is covariant $(p,q)-$five brane action \cite{Bergshoeff:2006gs} that
electrically couples to doublets of six forms that are dual to NSNS
and RR two forms.}. With the help of this argument we will be able
to find supergravity solutions corresponding to $(p,q)-$macroscopic
string and five branes that have correct NSNS and RR charges and
where the $(m,n)-$string probe has an expected properties. We show
that these solutions depend on two harmonic functions which is
different from the solutions found in
\cite{Schwarz:1995dk,Lu:1998vh} which however reflects the fact that
$(p,q)-$string can be considered as the bound state of
$p-$fundamental string and $q-$D1-branes even if this bound state is
not threshold and hence harmonic superposition rules cannot be
applied for it \cite{Tseytlin:1996bh}.

As the next step we extend this analysis to the case of NS5-brane
whose supergravity solution has been known for a long time
\cite{Callan:1991dj}. We perform $SL(2,Z)$ transformation of this
solution and find new solution corresponding to $(d,-b)$-five brane.
This solution is characterized by two harmonic functions whose
parameters depend on the charges of $(d,-b)-$five brane and on the
asymptotic values of moduli. We also analyze $(m,n)-$string probe
in this background and we show that it is equivalent to the dynamics
of $(m',n')$-string in original NS5-brane background where $m',n'$
are integers that again explain the puzzle found in
\cite{Kluson:2016pxg}.

Finally we consider $SL(2,Z)$ transformed solution of the a bound
state of $Q_5$ NS5-branes wrapped on four torus and $Q_1$
fundamental strings that are smeared over this four torus
\cite{Tseytlin:1996as}. We find solution that is characterized by
four harmonic functions that depend on the moduli of this solution
and charges with respect to RR and NSNS two forms.  This solution
has also an interesting  near horizon limit which is  $AdS_3\times
S^3$ with mixed three form fluxes with integer charges.
Integrability of superstring  in this background was studied
recently in
 \cite{Cagnazzo:2012se}. The main  idea of this
paper is to start with the pure RR background when the new WZ term
that represents the coupling to the NSNS flux is added. Using this
construction many new interesting results were derived
\cite{Hoare:2013pma,Hoare:2013ida,Hoare:2013lja,
Tseytlin:2015tpa,Lloyd:2014bsa,Sundin:2014ema,Kluson:2015lia}. It is
important to stress that in all these works the value of NSNS flux
can be interpreted as the deformation parameter that takes any real
value from the interval $(0,1)$. This is perfectly consistent from
the point of view of perturbative string theory  since classical
string does not couple to the dilaton. On the other hand this
approximation certainly breaks down  when we consider D1-brane in
this background that couples to the dilaton through
Dirac-Born-Infeld action. In order to analyze D1-brane in the
$AdS_3\times S^3$ with mixed three form fluxes we have to have
background with explicit values od dilaton and RR zero form too. It
is natural to presume that such a background arises as the near
horizon limit of the $SL(2,Z)-$transformed solution corresponding to
the bound state of NS-five branes and fundamental strings.
 We show that this is really true. More explicitly, we
show that $SL(2,Z)$-transformation and near horizon limit commutes
which is a generalization of the commutativity of S-duality and near
horizon limit
 found in \cite{Giveon:1998ns}. Then we study
$(m,n)-$string in $AdS_3\times S^3$ background with mixed three form
fluxes and  using the fact that the near horizon limit and
$SL(2,Z)-$transformation commutes we can map this $(m,n)-$string to
the $(m',n')-$string in $AdS_3\times S^3$ background with NSNS two
form flux. We show that for the special value of $(m,n)$ charges the
$(m,n)-$string in the $AdS_3\times S^3$ background is equivalent to
the fundamental string in $AdS_3\times S^3$ background with NSNS
three form flux  that can  be described by standard CFT techniques
\cite{Maldacena:1998bw,Giveon:1998ns,Maldacena:2000hw,Maldacena:2000kv}.
 Of course, this result
does not solve the problem of the analysis of fundamental string in
$AdS_3\times S^3$ with mixed fluxes which is very complicated and
deserves very special treatment \cite{Berkovits:1999im}.

This paper is organized as follows. In the next section
(\ref{second}) we review the basic fact about type IIB low energy
effective theory and suggest the main idea how to derive
$SL(2,Z)-$transformed solutions of type IIB supergravity equations
of motion. In section (\ref{third}) we apply this procedure to the
case of $(p,q)-$string background. In section (\ref{fourth}) we
perform the same analysis in case of $(p,q)-$five brane background
and we extend this analysis to the bound state of  NS5-branes and
fundamental strings in section  (\ref{fifth}). In section
(\ref{sixth}) we take the near horizon limit of this solution and
analyze its properties. Finally in conclusion (\ref{seventh}) we
outline our results and suggest possible extension of this work.

\section{$SL(2,Z)$ Covariance of Type IIB String
Theory}\label{second} In this section we review the basic facts about
  bosonic content
of the type IIB low energy effective action and we present general
idea how to find $SL(2,Z)$ transformed solution.

The type IIB theory has two three-form field strengths $H=dB,
F=dC^{(2)}$, where $H$ corresponds to NSNS three form while $F$
belongs to RR sector and does not couple to the usual string
world-sheet. Type IIB theory has also two scalar fields that can be
combined into a complex field $\tau= \chi+ie^{-\Phi}$. The dilaton
$\Phi$ is in the NSNS sector while $\chi$ belongs to the RR sector.
The other bose fields are the metric $g_{MN}$ in Einstein frame and
self-dual five form field strength $F_5$. However this field can be
consistently set to zero for solutions that we  study in this paper
and hence we do not include it to the action.
 Then it is possible to write down $SL(2,R)$
covariant form of the bosonic part of type IIB effective action
\begin{eqnarray}
S_{IIB}&=&\frac{1}{2\tkappa^2_{10}}\int d^{10}x
\sqrt{-g}(R+\frac{1}{4} \tr (\partial_M \mM
\partial^M\mM^{-1})-\frac{1}{12}\bH^T_{MNP} \mM
\bH^{MNP}) \ ,  \nonumber \\
2\tkappa^2_{10}&=&(2\pi)^7\alpha'^4
\ , \nonumber \\
\end{eqnarray}
where we have combined $B,C^{(2)}$ into
\begin{equation}
\bH=d\bB=\left(\begin{array}{cc} dB \\
dC^{(2)} \\ \end{array}\right)
\end{equation}
and where
\begin{eqnarray}
\mM=e^{\Phi}\left(\begin{array}{cc} \tau \tau^* & \chi \\
\chi & 1 \\ \end{array}\right)= e^{\Phi}\left(\begin{array}{cc}
\chi^2+e^{-2\Phi} & \chi \\
\chi & 1 \\ \end{array}\right) \ ,   \quad
 \det\mM=1 \ . \nonumber \\
\end{eqnarray}
This action has manifest invariance under the global $SL(2,R)$
transformation
\begin{equation}\label{trmB}
\hat{\mM}=\Lambda \mM \Lambda^T \ ,  \hat{\bB}=(\Lambda^T)^{-1} \bB
\ ,
\end{equation}
where
\begin{equation}
\Lambda=\left(\begin{array}{cc} a & b \\
c& d \\ \end{array}\right) \ , \quad
\det \Lambda=ad-bc=1 \ .
\end{equation}
Let us now introduce an action for $(m,n)-$string that couples
electrically to NSNS and RR two form and hence can be considered as
a source for corresponding fields. The action for $(m,n)$-string has
the form \footnote{For recent discussion, see
\cite{Kluson:2016pxg}.}
\begin{eqnarray}\label{mnaction}
S_{(p,q)}&=&-T_{D1}\int d\tau d\sigma(\sqrt{\bm^T\mM^{-1} \bm
}\sqrt{-\det g_{MN}\partial_\alpha x^M\partial_\beta x^N} +
\nonumber \\
&+&T_{D1}\int d\tau d\sigma \bm^T \bB_{MN}\partial_\tau
x^M\partial_\sigma x^N \ , \quad \bm=\left(\begin{array}{cc}
m \\
n \\ \end{array}\right) \ , \nonumber \\
\end{eqnarray}
where $T_{D1}=\frac{1}{2\pi\alpha'}$.
The action (\ref{mnaction})
 is invariant under global transformations (\ref{trmB}) on condition
 that $\bm$ transforms as
\begin{equation}
\hat{\bm}=\Lambda \bm \ .
\end{equation}
In this notation $m$ counts the number of fundamental strings while
$n$ counts the number of D1-branes  and hence they have to be
integer. This fact implies that   $\Lambda\in SL(2,Z)$. In other
words $SL(2,R)$ invariance of type IIB low energy effective action
is broken to its $SL(2,Z)$ subgroup due to the charge  quantization
condition. As a consequence the action that includes both type IIB
effective action and string probe action
\begin{equation}
S=S_{IIB}+S_{(p,q)}
\end{equation}
is invariant under $SL(2,Z)$ group rather than under $SL(2,R)$ group.
This fact will have a crucial consequence for the construction of $(p,q)-$
string and five brane solutions.

Using this basic presumption we now present the main
idea how to derive supergravity solution with
$(p,q)-$fundamental string as a source. Observe that
 we can write
\begin{equation}
\left(\begin{array}{cc} p \\
q \\ \end{array}\right)=
\left(\begin{array}{cc} p & b \\
 q & d \\ \end{array}\right)\left(\begin{array}{cc}
 1 \\
 0 \\ \end{array}\right) \Rightarrow \bm(p,q)=\Lambda(p,q)\bm_F \ , pd-bq=1 \
\end{equation}
and hence $(p,q)-$string action has the form
\begin{eqnarray}
S_{(p,q)}
&=&-T_{D1}\int d\tau d\sigma(\sqrt{\bm^T_F\tilde{\mM}^{-1} \bm_F
}\sqrt{-\det g_{MN}\partial_\alpha x^M\partial_\beta x^N} +
\nonumber \\
&+&T_{D1}\int d\tau d\sigma \bm^T_F \tilde{ \bB}_{MN}(p,q)
\partial_\tau x^M\partial_\sigma x^N \ ,  \nonumber \\
\end{eqnarray}
where
\begin{equation}
\tilde{\mM}=\Lambda^{-1}(p,q)\mM(p,q)(\Lambda^T(p,q))^{-1} \ , \quad
\tilde{\bB}= \Lambda^T(p,q)\bB(p,q) \ .
\end{equation}
Then with the manifest $SL(2,R)$ invariance of the Type IIB
effective action we find that it has the form
\begin{equation}
S_{IIB}=\frac{1}{2\tkappa^2_{10}}\int d^{10}x
\sqrt{-g}(R+\frac{1}{4} \tr (\partial_M \tilde{\mM}
\partial^M\tilde{\mM}^{-1})-\frac{1}{12}\tilde{\bH}^T_{MNP}
\tilde{\mM} \tilde{\bH}^{MNP}) \
\end{equation}
so that  $\tilde{\mM}$ and $\tilde{\bB}$ have the same functional
form as corresponding fields in case of fundamental string as the
source so that we denote its value with superscript $F$ and omit
tilde over them. In other words we find following components of
$\mM(p,q)$ and $\bB(p,q)$ corresponding to the $(p,q)-$string as a
source:
\begin{eqnarray}
\mM(p,q)=\Lambda(p,q)\mM_F\Lambda^T(p,q) \ ,
\bB(p,q)=(\Lambda^T)^{-1}(p,q)\bB_F \ . \nonumber \\
\end{eqnarray}
In this case the electric charge corresponding to this background
has the form
\begin{equation}
\mathbf{q}_{(p,q)}=\frac{1}{2\tkappa_{10}^2} \int_{S^8}\mM \star
\bH= \Lambda(p,q)\frac{1}{2\tkappa_{10}^2}\int_{S^8}\mM_F \star
\bH_F=
\left(\begin{array}{cc} p  \\
q \\ \end{array}\right)q_F \ ,
\end{equation}
where $q_F=\frac{1}{2\pi\alpha'}$ is the electric charge of the
fundamental string.

 In case of
solitaire $(p,q)-$brane we can argue in the similar way with the
difference  that this is a magnetic solution that is source free.
Further, the charge transforms in the same way as corresponding
field  strength
\begin{equation}\label{chargemaggen}
\mathbf{q}^5_{(d,-b)}=\frac{1}{2\tkappa_{10}^2}\int_{S^3}
\bH(p,q)= (\Lambda^T)^{-1}(p,q)
\frac{1}{2\tkappa_{10}^2}
\int_{S^3}\bH_F=
\left(\begin{array}{cc} d \\
-b \\ \end{array}\right)q_{NS5} \ ,
\end{equation}
where $q_{NS5}=\frac{1}{(2\pi)^5\alpha'^3}$ is the magnetic charge
of NS5-brane.
After the outline of this general procedure we proceed in next
sections to the explicit construction of the $SL(2,Z)-$transformed
solutions.


\section{$SL(2,Z)-$String solution}\label{third}
Let us start with the  fundamental string solution
\cite{Dabholkar:1990yf}
\begin{eqnarray}\label{Fsolution}
ds^2&=&G_{MN}^Fdx^M dx^N=\frac{1}{H_F}dx^2_{II}+dx^2_{\bot} \ ,
\quad
H_{m01}=\partial_m H^{-1}_F \ , \nonumber \\
e^{\Phi}&=&g_s\frac{1}{\sqrt{H_F}} \ , \quad  H_F=1+\frac{\alpha
g_s^2}{r^6} \ , \quad  \alpha=\frac{(2\pi)^6\alpha'^3}{6\Omega_7} \
, \quad B_{01}= \frac{1}{H_F}-1 \ ,
\nonumber \\
\end{eqnarray}
where $dx^2_{II}=-dt^2+dx_1^2 \ , dx^2_T=dx_m dx^m \ , m=2,\dots,9,
\  r^2=x_m x^m$ and  where the line element is expressed in string
frame. We use the notation when small $g_{MN}$ corresponds to the
Einstein frame metric while $G_{MN}$ corresponds to the string frame
metric. Note that   these two metrics  are related by rescaling
\begin{equation}
g_{MN}=e^{-\Phi/2}G_{MN} \ .
\end{equation}
Since $g_{MN}$ is invariant under
$SL(2,Z)$ transformation we derive relation between transformed and
original string frame metrics
\begin{equation}
\hat{G}_{MN}=e^{\frac{1}{2}(\hat{\Phi}-\Phi)}G_{MN} \ .
\end{equation}
Now we are ready to find solution that will be defined as $SL(2,Z)$
transformation  of the solution (\ref{Fsolution}) where the matrix
$\Lambda$ has the form
$\Lambda=\left(\begin{array}{cc} a & b \\
c & d \\ \end{array}\right) \ $
\begin{eqnarray}\label{rotFsolution}
\hat{\chi}&=&\frac{ac e^{-2\Phi}+bd}{c^2 e^{-2\Phi}+d^2} \ , \quad
e^{-\hat{\Phi}}=\frac{e^{-\Phi}}{c^2 e^{-2\Phi}+d^2} \ , \nonumber
\\
\hat{G}_{MN}&=&\sqrt{c^2 e^{-2\Phi}+d^2}G_{MN} \ , \nonumber \\
\hat{B}_{MN}&=&dB_{MN}
 \ , \quad \hat{C}^{(2)}_{MN}=-bB_{MN} \ , \quad  ad-bc=1 \ . \nonumber \\
\end{eqnarray}
As we argued in the previous section the new solutions have the
charges
\begin{equation}
\left(\begin{array}{cc} q^F_{NS}
\\
q^F_{RR} \\ \end{array}\right)=
\left(\begin{array}{cc} a  \\
c  \\ \end{array}\right)g_F \ .
\end{equation}
Now we write  explicitly $SL(2,Z)-$transformed line element
\begin{equation}
d\hat{s}^2=\frac{\sqrt{c^2+d^2g_s^2}}{g_s}
\sqrt{1+\frac{c^2}{c^2+d^2g_s^2}\frac{\alpha g_s^2}{r^6}}
(H^{-1}_Fd^2x_{II}+d^2x_{\bot}) \ .
\end{equation}
We see that it is natural to perform rescaling of the coordinates
\begin{equation}
\frac{(c^2+d^2g_s^2)^{1/4}}{\sqrt{g_s}}x^M=\hx^M \ , r^6=
\hr^6\frac{g_s^3}{(c^2+d^2g_s^2)^{3/2}} \ .
\end{equation}
On the other hand the solution found above still depends on the
string coupling $g_s$ of the original solution and non-physical
parameters $b,d$ that appear in $\Lambda$. However we would like to
express the new solution using the assymptotic values of
$\hat{\Phi}$ and $\hat{\chi}$ together with $a$ and $c$ that correspond
to $q^F_{NS}$ and $q_{NS}^R$ charges. In order to do
this we take the limit $r\rightarrow \infty$ in
$e^{-\hat{\Phi}}$ and $\hat{\chi}$ given in (\ref{rotFsolution}) and
 we obtain
\begin{eqnarray}
\lim_{r\rightarrow \infty}e^{-\hat{\Phi}}=\frac{1}{\hat{g}_s}=
\frac{g_s}{c^2 +d^2g_s^2} \ , \nonumber \\
\lim_{r\rightarrow \infty}\hat{\chi}\equiv \chi_0=\frac{ac+bd
g_s^2}{
c^2+d^2 g_s^2} \ . \nonumber \\
\end{eqnarray}
If we multiply the last equation with $c$ and use the fact that
$ad-bc=1$ we obtain
\begin{equation}
g_s d=\hg_s (a-c\chi_0) \ , \quad  g_s=\frac{1}{\hg_s}
(c^2+\hg_s^2(a-c\chi_0)^2) \ .
\end{equation}
Then we define two harmonic functions
\begin{eqnarray}
\hat{H}&\equiv & H_F(\hat{r})=1+\frac{\alpha
\hg_s\sqrt{c^2+\hg_s^2(a-c\chi_0)^2}} { \hr^6} \ ,
\nonumber \\
 \hat{H}'
&=& 1+\frac{c^2}{\sqrt{c^2+\hg_s^2
(a-c\chi_0)^2}}\frac{\alpha \hg_s}{\hr^6} \  \nonumber \\
\end{eqnarray}
so that the line element has the final form
\begin{equation}
d\hat{s}^2=\sqrt{\hat{H}'} (\hH^{-1}d^2\hx_{II}+d^2\hx_{\bot}) \ .
\end{equation}
As a check note that for $a=\chi_0=0$ we obtain that
$\hat{H}=\hat{H}'\equiv H_{(0,1)}$ and hence
\begin{equation}
d\hat{s}^2=\frac{1}{\sqrt{H_{(0,1)}}}d^2\hx_{II}+\sqrt{H_{(0,1)}}
d^2\hx_{\bot} \ , \quad H_{(0,1)}=1+\frac{\alpha \hg_s}{\hr^6}
\end{equation}
which corresponds to the line element of the D1-brane which is
S-dual to the fundamental string solution.

 Finally we express dilaton as a function of $\hH$ and $\hH'$
\begin{equation}
e^{-\hat{\Phi}}=\frac{1}{\hg_s}\frac{\sqrt{\hat{H}}}{\hat{H}'} \
\end{equation}
and find components of NSNS and RR two forms in the
new coordinates $\hx$. To do this we use the fact that
\begin{eqnarray}
\hat{B}=dB_{MN}dx^M \wedge dx^N=
\frac{\hg_s(a-c\chi_0)}{\sqrt{c^2+\hg_s^2(a-c\chi_0)^2}}
B_{MN}d\hx^M\wedge d\hx^M
\end{eqnarray}
and consequently
\begin{equation}
\hB_{01}=\frac{\hg_s(a-c\chi_0)}{\sqrt{c^2+\hg_s^2(a-c\chi_0)^2}}\left(
\frac{1}{\hH(\hx)}-1\right) \ .
\end{equation}
In the same way we find
\begin{equation}
\hat{C}_{01}=\frac{c-\hg_s^2(a-c\chi_0)\chi_0}
{\hg_s\sqrt{c^2+\hg_s^2(a-c\chi_0)^2}} \left(
\frac{1}{\hH(\hx)}-1\right) \nonumber \\
\end{equation}
using $b=\frac{ad-1}{c}$. In summary, we claim that the supergravity
solution corresponding to $(a,c)-$string has the form
\begin{eqnarray}\label{stringfinal}
 d\hat{s}^2&=&\sqrt{\hat{H}'}
(\hH^{-1}d^2\hx_{II}+d^2\hx_{\bot}) \  , \quad
e^{-\hat{\Phi}}=\frac{1}{\hg_s}\frac{\sqrt{\hat{H}}}{\hat{H}'} \ ,
\nonumber \\
\hB_{01}&=&\frac{\hg_s(a-c\chi_0)}{\sqrt{c^2+\hg_s^2(a-c\chi_0)^2}}\left(
\frac{1}{\hH(\hx)}-1\right) \ , \quad
\hat{C}_{01}=\frac{c-\hg_s^2(a-c\chi_0)\chi_0}
{\hg_s\sqrt{c^2+\hg_s^2(a-c\chi_0)^2}} \left(
\frac{1}{\hH(\hx)}-1\right) \nonumber \\
\hat{H}&=&1+\frac{\alpha \hg_s\sqrt{c^2+\hg_s^2(a-c\chi_0)^2}} {
\hr^6} \ , \quad
 \hat{H}'
= 1+\frac{c^2}{\sqrt{c^2+\hg_s^2
(a-c\chi_0)^2}}\frac{\alpha \hg_s}{\hr^6} \ . \nonumber \\
\end{eqnarray}
It is instructive to compare this solution with $(a,c)$-fundamental
string solution found in \cite{Schwarz:1995dk}. The main  difference
is that our solution depends on two harmonic functions as opposite
to the solution derived in \cite{Schwarz:1995dk}. In some way this
is a reflection of the fact that we have a  bound state of D1-brane
and fundamental string even if this superposition does not
correspond to the harmonic superposition rule \cite{Tseytlin:1996bh}
 as this bound state is not marginal. Further, the arguments of the
 harmonic functions are different from the expression used in
\cite{Schwarz:1995dk} which however implies that our solution is
defined with the help of $SL(2,Z)$ matrix rather than $SL(2,R)$
matrix that was used in \cite{Schwarz:1995dk}. As a consequence the
solution (\ref{stringfinal}) behaves consistently from the  probe
$(m,n)-$string point of view. To see this explicitly let us consider
 probe
$(m,n)-$string in this background when the action has the form
 \begin{eqnarray}
S_{(m,n)}&=&-T_{D1}\int d\tau d\sigma(\sqrt{\bm^T\hat{\mM}^{-1} \bm
}\sqrt{-\det \hat{g}_{MN}\partial_\alpha \hx^M\partial_\beta \hx^N} +
\nonumber \\
&+&T_{D1}\int d\tau d\sigma \bm^T \hat{\bB}_{MN}\partial_\tau
\hx^M\partial_\sigma \hx^N \ ,  \nonumber \\
\end{eqnarray}
where $\hg_{MN}$ is the Einstein frame metric in rescaled
coordinates
\begin{equation}
 \hg_{MN} =
\frac{\hg_s}{\sqrt{c^2+\hg_s^2(a-c\chi_0)^2}}g_{MN}(\hr)
\ .
\end{equation}
However due to the fact that the pullback of the Einstein metric and
two forms is invariant under rescaling by definition we can easily
use the original variables $x$ instead of $\hx$. Then the probe
action has the form
 \begin{eqnarray}
S_{(m,n)}
&=&-T_{D1}\int d\tau d\sigma(\sqrt{m'^2+n'^2
e^{-2\Phi_F}}\sqrt{-\det G^F_{MN}\partial_\alpha x^M\partial_\beta
x^N} +
\nonumber \\
&+&T_{D1}\int d\tau d\sigma m'B^F_{MN}\partial_\tau
x^M\partial_\sigma x^N \ , \nonumber \\
\end{eqnarray}
where
\begin{equation}\label{bmF'}
\bm'=\left(\begin{array}{cc} m '\\
n' \\ \end{array}\right)=\left(\begin{array}{cc} dm-bn \\
-cm+an \\ \end{array}\right) \ .
\end{equation}
From the previous action we see that the problem of the analysis of
the dynamics of $(m,n)-$string in $(a,c)-$string background is
reduced to the analysis of $(m',n')-$string in the fundamental
string background where $m',n'$ are evaluated at (\ref{bmF'}). The
beautiful analysis of this problem was performed in \cite{Bak:2004tp}
and we will not reviewed it here. We also see that for $m=a,n=c$ we
obtain that $m'=1,n'=0$ which is again consistent with the picture
of probe F-string in fundamental string background that is rotated
by $SL(2,Z)$ transformation. In other words our solution solves the
issue that was found in our previous paper \cite{Kluson:2016pxg}.

\section{$(d,-b)-$Five Brane Solution}\label{fourth}
In this section we find $(d,-b)$-five brane solution when we perform
$SL(2,Z)$ transformation of NS5-brane supergravity solution. Recall
that this  solution has the form \cite{Callan:1991dj}
\begin{eqnarray}\label{NS5back}
ds^2&=&G_{MN}^{NS5}dx^M dx^N=dx^2_{II}+H_{NS5}dx^2_{\bot} \ , \quad
e^{\Phi}=g_s
H_{NS5}^{1/2} \ , \nonumber \\
 H_{NS5}&=&1+\frac{\alpha'}{r^2} \ , \quad
H_{mnp}^{NS5}=\epsilon_{mnpq}\partial_q
H_{NS5}\ , \nonumber \\
\end{eqnarray}
where $dx_{II}^2=\eta_{\mu\nu}dx^\mu dx^\nu \ , \mu,\nu=0,\dots,5, \ 
dx^2_{\bot}=dx_m dx^m \ , m=6,\dots,9 \ , r^2=x_m x^m$.
As in previous section we perform
 $SL(2,Z)$ transformations
\begin{eqnarray}
\hat{\chi}&=&\frac{ac e^{-2\Phi}+bd}{c^2 e^{-2\Phi}+d^2} \ , \quad
e^{-\hat{\Phi}}=\frac{e^{-\Phi}}{c^2 e^{-2\Phi}+d^2} \ , \nonumber
\\
\hat{G}_{MN}&=&\sqrt{c^2 e^{-2\Phi}+d^2}G^{NS5}_{MN} \ , \quad
\hat{B}_{MN}=dB^{NS5}_{MN}
 \ , \quad \hat{C}^{(2)}_{MN}=-bB^{NS5}_{MN} \ .\nonumber \\
\end{eqnarray}
We will argue below that $d$ and $b$ are proportional to $NSNS$ and $RR$ magnetic
charges of five brane. For that reason 
 we would like to express  transformed solution as a function of $d,b$
together with the string coupling
constant
 $\hg_s$ and assymptotic values of  $\hat{\chi}_{(0)}$.

To begin with we  note  that  the transformed line element has the
form
\begin{equation}
d\hat{s}^2=\sqrt{c^2+d^2g_s^2H_{NS5}}\frac{1}{g_s}(H^{-1/2}_{NS5}
d^2x_{II}+H^{1/2}_{NS5}d^2x_{\bot}) \ .
\end{equation}
Now we observe that we can write
\begin{equation}
\sqrt{c^2+d^2g_s^2H_{NS5}}=\sqrt{c^2+d^2g_s^2}\sqrt{1+\frac{d^2}{c^2+d^2g_s^2}\frac{
 N \alpha'g_s^2}{r^2}}=\sqrt{c^2+d^2g_s^2}\hH'^{1/2} \ .
\end{equation}
For $r\rightarrow\infty$ we obtain
\begin{eqnarray}
& &\lim_{r\rightarrow \infty}e^{-\hat{\Phi}}=\frac{1}{\hat{g}_s}=
\frac{g_s}{c^2 +d^2g_s^2} \ , \nonumber \\
& &\lim_{r\rightarrow \infty}\chi \equiv \chi_0=\frac{ac+bd g_s^2}{
c^2+d^2 g_s^2} \  . \nonumber \\
\end{eqnarray}
Now if we multiply the first expression
with $d$ and use the fact that
$ad-bc=1$ we can express $c$ in terms of $\chi_0,\hat{g}_s,d,b$ as
\begin{equation}
c=\hat{g}_s g_s(d\chi_0-b)
\end{equation}
and consequently
\begin{eqnarray}
 g_s=\frac{\hg_s}{\hg_s^2(d\chi_0-b)^2+d^2} \ .
 \nonumber \\
 \end{eqnarray}
 We further
 rescale  coordinates as
\begin{equation}
(\hg_s^2(d\chi_0-b)^2+d^2)^{1/4}x^M=\hx^M \Rightarrow
r^2=\frac{\hr^2}{\sqrt{\hg_s^2(d\chi_0-b)^2+d^2}}
\end{equation}
and hence $\hH$ and $\hH'$ have the form
\begin{eqnarray}
\hH&=&1+\frac{ N\alpha'}{\hr^2} \sqrt{\hg_s^2(d\chi_0-b)^2+d^2} \ ,
\nonumber \\
\hH'&=& 1+\frac{ N\alpha'} {\hr^2}\frac{d^2}{ \sqrt{\hg_s^2
(d\chi_0-b)^2+d^2}}  \ . \nonumber \\
\end{eqnarray}
Finally we  determine  components of RR and NSNS two
forms. It is useful to express them in covariant independent
formulation and we obtain
\begin{eqnarray}
\hat{\bH}=
2\left(\begin{array}{cc}
d \\
-b \\ \end{array}\right) \alpha'\epsilon_3 \ , \nonumber \\
\end{eqnarray}
where $\epsilon_3$ is the volume of three sphere.
As a check let us calculate NSNS magnetic charge  of $( d,-b)-$five brane
\begin{equation}
q^5_{NS}=\frac{1}{2\tkappa_{10}^2}\int_{S^3} \hat{H}=
d q_{NS5} \ .
\end{equation}
In the same way we determine RR charge
\begin{equation}
q^5_{RR}=-bq_{NS5}\
\end{equation}
which is in agreement with the general result 
(\ref{chargemaggen}). Finally we determine 
 the space-time dependence of the dilaton
\begin{equation}
e^{-\hat{\Phi}}=
\frac{1}{\hg_s}\frac{\sqrt{\hH}}{\hH'} \ .
\end{equation}
Now we proceed to the analysis of the  probe $(m,n)-$string
 in this background. In the
same way as in previous section we find that the action has the form
\begin{eqnarray}\label{mnNS5action}
S_{(m,n)}&=&-T_{D1}\int d\tau d\sigma(\sqrt{m'^2 +n'^2
e^{-2\Phi_{NS5}}}\sqrt{-\det G^{NS5}_{MN}\partial_{\alpha}
x^M\partial_{\beta} x^N} +
\nonumber \\
&+&T_{D1}\int d\tau d\sigma m'B^{NS5}_{MN}\partial_{\tau}
x^M\partial_{\sigma} x^N \  , \nonumber \\
\end{eqnarray}
where
\begin{equation}
\bm'=\left(\begin{array}{cc} d & -b \\
-c & a \\ \end{array}\right)\left(\begin{array}{cc} m \\
n \\ \end{array}\right)= \left(\begin{array}{cc} dm-bn \\
-cm+an \\ \end{array}\right) \ .
\end{equation}
As we could expect the $(m,n)-$action in $(d,-b)-$five brane background
is equivalent to the  action of $(m',n')$ string in the background
of $NS5-$brane. The most interesting case occurs for
\begin{equation}
m=a \ , \quad n=c \
\end{equation}
that implies $m'=1 \ , n'=0$ and hence the action corresponds to the
fundamental string in NS5-brane action. In order to analyze main
properties of this configuration we consider string stretched along
$x^0,x^1$ directions, impose the static gauge $x^0=\tau ,
x^1=\sigma$ and finally consider time dependent radial coordinate
only. Then the induced metric has the form
\begin{equation}
g_{\tau\tau}=\frac{1}{\sqrt{g_s}} (-H^{-1/4}_{NS5}+H^{3/4}_{NS5}\dot{R}^2) \ ,
\quad g_{\sigma\sigma}=\frac{1}{\sqrt{g_s}}H^{-1/4} \ , \dot{R}=\frac{dR}{dt}
\end{equation}
and hence the  action  (\ref{mnNS5action}) has the form
\begin{equation}\label{RNS5}
S_{(a,c)}=-\frac{1}{2\pi\alpha'}\int d\tau d\sigma
\sqrt{1-H_{NS5}\dot{R}^2} \ .
\end{equation}
From (\ref{RNS5}) we see that
 there is no potential for  the fundamental string probe in
the NS5-brane background which nicely demonstrates the fact that
fundamental string together with NS5-brane background can form
marginal bound state. Expressing in original variables there is no
potential for $(a,c)$-string in $(d,-b)$-five brane background when
$ad-bc=1$.

Another interesting case occurs for
\begin{equation}
m=b \ , \quad  n=d
\end{equation}
that implies $m'=0 , n'=1$. In other words the dynamics of
$(b,d)-$string in $(d,-b)$-five brane background is equivalent to
the motion of D1-brane in NS5-brane background. The dynamics of this
configuration was analyzed in \cite{Kutasov:2004dj} and we will not
repeat it here. Finally note that the motion of the general
$(m',n')-$string in NS5-brane background is simple generalization of
the case of the electrified brane
\cite{Chen:2004vw,Nakayama:2004ge}. Since this generalization is
trivial we will not repeat it here and recommend the original papers
for more details.

\section{$SL(2,Z)$ Transformation  of NS5-brane and F-String Bound
State}\label{fifth} In this section we perform $SL(2,Z)$
transformation of the  supergravity solution with $Q_1$ fundamental
strings and $Q_5$-five branes that has the form
\cite{Tseytlin:1996as}
\begin{eqnarray}\label{NS5F1back}
e^{-2\Phi}&=&\frac{1}{g^2_s}f_5^{-1}f_1 \ , \quad
B_{05}=\frac{1}{f_1}-1 \ , \quad H_{mnp}= \epsilon_{mnpq}\partial_q
f_5 \ , m,n,p,q=1,2,3,4 \ ,
\nonumber \\
ds^2&=& f_{1}^{-1}(-dt^2+dx^2_5)+f_5(dx_1^2+\dots+
dx^2_4)+(dx_6^2+\dots +dx^2_9) \ ,
\nonumber \\
\end{eqnarray}
and where
\begin{equation}
f_1=1+\frac{r_1^2}{r^2}=1+\frac{16 \pi^4 g_s^2 \alpha'^3
Q_1}{V_4r^2} \ , \quad f_5=1+\frac{r^2_5}{r^2}=1+\frac{\alpha'
Q_5}{r^2} \  ,
\end{equation}
where $r^2=\sum_{m=1}^4 x_m^2$, where $x_m$ are coordinates in the
space transverse to NS5-branes wrapped over four torus
with volume $V_4=(2\pi)^4\alpha'^2v$. This
fact implies that each $x^i,i=6,\dots,9$ are identified with period
$2\pi v^{1/4}\alpha'^{1/2}$. Note also  that the
 fundamental strings are
smeared over this  four torus.

Let us now perform  $SL(2,Z)$ transformation  of the background
(\ref{NS5F1back})
and we find that  the charges corresponding to the fundamental strings and
NS5-branes  are equal to
 \begin{equation}
\left(\begin{array}{cc} q^F_{NS}
\\
q^F_{RR} \\ \end{array}\right)=
\left(\begin{array}{cc} a  \\
c  \\ \end{array}\right)Q_1q_F \ , \quad
\left(\begin{array}{cc} q^5_{NS} \\
q^5_{RR} \\ \end{array}\right)= \left(\begin{array}{cc} d

\\
-b \\ \end{array}\right)Q_5q_{NS5} \
\end{equation}
while the line element has the form
\begin{eqnarray}
d\hat{s}^2
&=&\frac{\sqrt{c^2+d^2g_s^2}}{g_s} \sqrt{1+\frac{c^2r_1^2+d^2 g_s^2
r_5^2}{c^2+d^2g_s^2}\frac{1}{r^2}} \times \nonumber \\
&\times & \left(\frac{1}{f_1\sqrt{f_5}}(-dt^2+dx^2_5)+
\sqrt{f_5}(dx_1^2+\dots + dx^2_4)+ \frac{1}{\sqrt{f_5}} (dx^2_6+\dots +
dx_9^2)\right) \ .  \nonumber \\
\end{eqnarray}
We see that it is again natural to perform rescaling
\begin{equation}
\frac{(c^2+d^2g_s^2)^{1/4}}{\sqrt{g_s}}x^M=\hx^M \ , \quad  r^2=
\hr^2\frac{g_s}{\sqrt{c^2+d^2g_s^2}} \ . 
\end{equation}
Further, in the limit  $r\rightarrow\infty$ we have
\begin{eqnarray}
\frac{1}{\hat{g}_s}=
\frac{g_s}{c^2 +d^2g_s^2} \ , \quad 
 \chi_0=\frac{ac+bd g_s^2}{
c^2+d^2 g_s^2} \ \nonumber \\
\end{eqnarray}
and we again want to 
 express $f_1$ and $f_5$ as functions of $\hg_s$ and $\chi_0$ 
 together with   the numbers that are
proportional to the corresponding charges.
 If we proceed in the same way as in previous two
sections we find
%
\begin{equation}
\hf_1=1+\frac{16\pi^4\alpha^3 \hg_s Q_1}{\hat{V}_4 \hr^2}
\sqrt{c^2+\hg_s^2(a-c\chi_0)^2} \ ,
\end{equation}
where we also used the fact  that under rescaling given above the
coordinates $\hx_i,i=6,\dots,9$ have identifications $\frac{2\pi
v^{1/4}\alpha'(c^2+d^2g_s^2)^{1/4}}{\sqrt{g_s}}$ and hence
\begin{equation}
V_4=\hat{V}_4
\frac{c^2+\hg_s^2(a-c\chi_0)^2}{\hg_s^2} \ .
\end{equation}
In case of $f_5$ we proceed as in section (\ref{third}) and we
obtain
\begin{equation}
f_5=1+\frac{\alpha'Q_5\sqrt{\hg_s^2(d \chi_0-b)^2+d^2}}{\hr^2} \ .
\end{equation}
Finally we write
\begin{eqnarray}
1+\frac{c^2r_1^2+d^2 g_s^2
r_5^2}{c^2+d^2g_s^2}\frac{1}{r^2}
=\hf'_1+\hf'_5-1 \nonumber \\
\hf'_1=1+\frac{16\pi^4\alpha'^3 Q_1 c^2 \hg_s}{\hr^2 \hat{V}_4
\sqrt{c^2+\hg_s^2(a-c\chi_0)^2}} \ , \nonumber \\
f'_5=1+\frac{\alpha' Q_5}{\hr^2} \frac{d^2}{ \sqrt{d^2+\hg_s^2
(d\chi_0-b)^2}} \ . \nonumber \\
\end{eqnarray}
Collecting all these results together we obtain the line element in
the form
\begin{equation}\label{F1NSfinal}
d\hat{s}^2=\sqrt{\hf'_1+\hf'_5-1}\left( \frac{1}{\hf_1\sqrt{\hf_5}}
(-d\hx_0^2+d\hx^2_5)+\sqrt{\hf_5}(d\hx_1^2+\dots+
d\hx_4^2)+\frac{1}{\sqrt{\hf_5}}(d\hx^2_6+\dots + d\hx^2_9)\right)
\end{equation}
while the dilaton and RR zero form are equal to
\begin{eqnarray}\label{dilfinalNS}
e^{-\hat{\Phi}}&=&
\frac{1}{\hg_s}\frac{\sqrt{\hf_1\hf_5}}{\hf'_1+\hf'_5-1} \ ,
\nonumber \\
\hat{\chi}&=&\frac{1}{c^2+\hg_s^2(a-c\chi_0)^2} \frac{ac\hf_1+
\frac{bd}{\hg_s^2}(c^2+\hg_s^2(a-c\chi_0)^2)\hf_5 }{\hf'_1+\hf'_5-1}
\nonumber \\
\end{eqnarray}
Finally the non-zero RR and NSNS two and three forms have the form
\begin{eqnarray}\label{C05final}
\hat{C}_{05}&=&-\frac{b}{\sqrt{c^2+\hg_s^2(a-c\chi_0)^2}}\left(\frac{1}{\hf_1}-1\right)
\ , \nonumber \\
\hat{B}_{05}&=&\frac{d}{\sqrt{c^2+\hg_s^2(a-c\chi_0)^2}}\left(\frac{1}{\hf_1}-1\right)
\ . \nonumber \\
\end{eqnarray}
For the contribution from $5-$branes we obtain
\begin{eqnarray}\label{HFNS5final}
\hH_{mnp}=2dQ_5 \alpha'(\epsilon_{S^3})_{mnp} \ , \quad
\hF_{mnp}=-2bQ_5 \alpha'(\epsilon_{S^3})_{mnp} \ . \nonumber \\
\end{eqnarray}
As the check of the validity of our solution let us consider S-duality
transformation when $a=b=0$ and $c=1, b=-1$ that also implies
$\chi_0=0$. Then it is easy to see that the  harmonic functions defined
above have the form
\begin{eqnarray}
\hf_1=\hf'_1=1+\frac{16\pi^4 \alpha'^3 \hg_s Q_1 }{\hat{V}_4
\hat{r}^2} \ , \quad \hf_5=1+\frac{\alpha'Q_5\hg_s}{\hat{r}^2} \ , \quad
\hf'_5=1
\nonumber \\
\end{eqnarray}
and hence the line element has the form
\begin{equation}
d\hat{s}^2= \frac{1}{\sqrt{\hf_1\hf_5}}
(-d\hx_0^2+d\hx^2_5)+\sqrt{\hf_1\hf_5}(d\hx_1^2+\dots+
d\hx_4^2)+\frac{\sqrt{\hf_1}}{\sqrt{\hf_5}}(d\hx^2_6+\dots +d\hx^2_9)
\end{equation}
and dilaton
\begin{eqnarray}
e^{-\hat{\Phi}}=\frac{1}{\hg_s}\sqrt{\frac{\hf_5}{\hf_1}} \
\end{eqnarray}
which is precisely the D1-D5-brane background
\cite{Maldacena:1998bw}, for nice review see \cite{David:2002wn}.

We are mainly interested in the near horizon limit of the background
specified by the equations (\ref{F1NSfinal}),(\ref{dilfinalNS}),
(\ref{C05final}) and (\ref{HFNS5final}) since we expect that it
leads to $AdS_3\times S^3$ background with mixed fluxes.
 We do it in the next section.

\section{Near Horizon Limit}\label{sixth}
We consider two ways how to perform the near horizon
limit of $SL(2,Z)$ transformed  solution of
  the bound state of $NS-$five branes and fundamental strings.
 In the first case we firstly
take the near horizon limit and then perform the $SL(2,Z)$
transformation. In the second one we consider opposite situation when we
firstly perform the $SL(2,Z)$ transformation
 and then we take the near horizon limit.
Finally we compare these two results.

In the first case we start with  the solution (\ref{NS5F1back})
and take its near horizon limit $r\ll r_{1,5}$ and  we obtain
 the background in the form
\begin{eqnarray}
ds^2=\frac{r^2}{r^2_1}(-dt^2+dx^2_5)+\frac{r_5^2}{r^2} (dx_1^2+
\dots +dx^2_4)+(dx_6^2+\dots +dx^2_9) \ .  \nonumber \\
\end{eqnarray}
Let us rescale the time and $x_5$ coordinate as
\begin{equation}
\frac{r_5}{r_1}dr=d\hat{t} \ , \quad \frac{r_5}{r_1}dx_5=d\hat{x}_5
\end{equation}
so that the line element has the form
\begin{equation}
ds^2=\frac{r^2}{L^2}(-d\hat{t}^2+d\hat{x}_5^2)+\frac{L^2}{r^2}dr^2+
L^2d\Omega_3+ (dx_6^2+\dots dx_9^2) \ , L^2=r_5^2 \
\end{equation}
while the dilaton is constant and equal to
\begin{equation}
e^{-2\Phi}=\frac{1}{g_s^2}\frac{r_1^2}{r_5^2} \ .
\end{equation}
Note that thanks to the rescaling we have
\begin{equation}
H_{r05}=2\frac{r}{L^2} \
\end{equation}
so that the form $H$ can be written using the volume
element of $AdS_3$ as
\begin{equation}
H=\frac{2}{L}\epsilon_{AdS_3} \ ,
\epsilon_{AdS_3}=\sqrt{\det g_{AdS_3}}dr\wedge d\hat{t}\wedge
d\hat{x}_5 \ .
\end{equation}
The flux from the NS5-brane source has the form
\begin{eqnarray}
H=
2L^2 \epsilon_{S^3} \ . \nonumber \\
\end{eqnarray}
Finally we  introduce dimensionless coordinates on $AdS_3$ in the following way
\begin{equation}
\hat{t}=L\tilde{t} \ , \quad \hat{x}_5=L \tilde{x}_5
\end{equation}
so that $
\epsilon_{AdS_3}=
L^3 \tilde{\epsilon}_{AdS_3}$
and hence   we obtain the  result
\begin{equation}\label{HNSAdS}
H=2L^2 (\tilde{\epsilon}_{AdS_3}+\epsilon_{S^3}) \
\end{equation}
and the line element in the form
\begin{eqnarray}\label{lineAdSNS}
ds^2&=&L^2\left[\tilde{r}^2(-
d\tilde{t}^2+d\tx^2)+\frac{d\tilde{r}^2}{\tilde{r}^2}+d\Omega_3\right]+ds_T^2=
\nonumber \\
&=&L^2[ds^2_{\widetilde{AdS}_3}+ds^2_{\Omega^3}]+ds_T^2 \ , \nonumber \\
\end{eqnarray}
where $ds^2_T=dx_6^2+\dots +dx_9^2$ and where $ds^2_{\tilde{AdS}_3}$ is the line
element of $AdS_3$ space expressed using  dimensionless variables.

Since the solution given above
 is a consistent solution of type IIB supergravity it is possible
 to perform $SL(2,Z)$ transformation of this solution.
 As a result we obtain
\begin{equation}
d\tilde{s}^2=\sqrt{\frac{c^2}{g_s^2}\frac{r_1^2}{r_5^2}+d^2}
\left[L^2[ds^2_{\widetilde{AdS}_3}+ds^2_{\Omega^3}]+ds_T^2\right] \ .
\end{equation}
 We see that the new solution has
the curvature radius $\tilde{L}^2=\sqrt{\frac{c^2}{g_s^2}\frac{r_1^2}{r_5^2}+d^2}L^2=
\frac{1}{g_s}\sqrt{c^2 r_1^2+d^2g_s^2 r^2_5}r_5$.
Further, there are following NSNS and RR three forms
\begin{eqnarray}
\tilde{H}=
2d\alpha'Q_5
(\tilde{\epsilon}_{AdS_3}+\epsilon_{S^3}) \ ,
 \quad
\tilde{F}=
-2b\alpha'Q_5(\tilde{\epsilon}_{AdS_3}+\epsilon_{S^3})
\nonumber \\
\end{eqnarray}
and dilaton and zero RR form are equal to
\begin{eqnarray}
e^{-\tilde{\Phi}}\equiv \frac{1}{\hg_s}=
\frac{\sqrt{Q_1 Q_5v}}{c^2 Q_1+d^2 Q_5v}
 \ , \quad
\tilde{\chi}=
\frac{ac Q_1+bd vQ_5}{c^2 Q_1+d^2 Q_5 v}
 \ . \nonumber \\
\end{eqnarray}
Let us now consider the case  when we firstly perform $SL(2,Z)$
duality transformation and then take the near horizon limit.
Explicitly, we begin with the $SL(2,Z)$ transformed background
\begin{eqnarray}
d\hat{s}^2
&=&\frac{1}{g_s}\sqrt{c^2r_1^2+d^2 g_s^2 r_5^2}\left( \frac{r^2}{r_1^2
r_5}(-dt^2+dx_5^2)+\frac{r_5}{r^2}dr^2+ r_5 d\Omega_3+
\frac{1}{r_5}(dx_6^2+\dots +dx_9^2)\right) \ .  \nonumber \\
\end{eqnarray}
We rescale $t$ and $x^5$ coordinates as
\begin{eqnarray}
t\frac{1}{g_sr_1}\sqrt{c_1^2r_1^2+d^2 g_s^2 r_5^2}=\hat{t} \ ,
\quad x^5\frac{1}{g_s}\sqrt{c_1^2r_1^2+d^2 g_s^2 r_5^2}=\hx^5 \ ,
\nonumber \\
\end{eqnarray}
so that the line element has the form
\begin{eqnarray}
d\hat{s}^2&=&r^2\frac{g_s}{r_5\sqrt{c^2 r_1^2+d^2g_s^2 r_5^2}}
(-d\hat{t}^2+d\hat{x}_5^2)+ \frac{1}{g_s}\sqrt{c^2r_1^2+d^2 g_s^2
r_5^2}r_5 \frac{dr^2}{r^2}+ \frac{1}{g_s}\sqrt{c^2r_1^2+d^2 g_s^2
r_5^2}r_5
d\Omega_3^2+ \nonumber \\
&+&\frac{1}{g_s r_5}\sqrt{c^2r_1^2+d^2 g_s^2 r_5^2}
(dx^2_6+\dots dx^2_9) \ ,  \nonumber \\
\end{eqnarray}
where the expression on the first line corresponds to
the line element of  $AdS_3\times S^3$ with the
curvature radius
\begin{equation}
\hat{L}^2=\frac{r_5\sqrt{c^2 r_1^2+d^2g_s^2 r_5^2}}{g_s}=\alpha'
\sqrt{\frac{c^2 Q_1 Q_5+d^2 Q_5^2 v}{v}} \ 
\end{equation}
and we see that $\hat{L}$ and $\tilde{L}$ coincide.
 In the same way we obtain
\begin{eqnarray}
e^{-\hat{\Phi}}
=\frac{\sqrt{Q_1 Q_5v}}{c^2 Q_1+d^2 Q_5v}
 \ ,  \quad
 \hat{\chi}=
\frac{ac Q_1+bd vQ_5}{c^2 Q_1+d^2 Q_5 v} \  \nonumber \\
\end{eqnarray}
and we again see that these expressions coincide 
 with $\tilde{\Phi}$ and $\tilde{\chi}$. 
Now we focus on  the near horizon limit of forms. 
In case of  $\hat{B}_{05}$  we obtain
\begin{eqnarray}
\hat{B}_{05}=
 \frac{g^2_s } {c_1^2r_1^2+d^2 g_s^2
r_5^2}r^2  \nonumber \\
\end{eqnarray}
so that using the rescaled coordinates
\begin{equation}
\hat{t}=\hat{L}\tilde{t} \ , \quad  \hat{r}=\hat{L}\tilde{r} \ , \quad  \hat{x}_5=\hat{L}\tilde{x}_5
\end{equation}
we obtain
\begin{equation}
\hat{H}=
2dQ_5\alpha'
 (\epsilon_{\widetilde{AdS}_3}
+\epsilon_{S^3}) \  , \quad
\hat{F}=
-2b Q_5\alpha'
(\epsilon_{\widetilde{AdS}_3} +\epsilon_{S^3})
 \
\end{equation}
that agree with $\tilde{H}$ and $\tilde{F}$. In summary
we showed an important result that  the near horizon
limit and $SL(2,Z)$ transformation commutes which implies
 that the $AdS_3\times S^3$ background with mixed fluxes
can be derived from $AdS_3\times S^3$ background through $SL(2,Z)$
rotation.

We will now going to analyze consequences of this result for the
dynamics of the probe string in this background. Using the same
arguments as in previous section we find the action in the form
\begin{eqnarray}
S
&=&-T_{D1}\int d\tau d\sigma \sqrt{m'^2+n'^2e^{-2\Phi_{NS}}}
\sqrt{-\det (G^{NS}_{MN}\partial_\alpha x^M\partial_\beta x^N)}+
\nonumber \\
&+&T_{D1}\int d\tau d\sigma m'B^{NS}_{MN}\partial_\tau
x^M\partial_\sigma x^N \ ,
\nonumber \\
\end{eqnarray}
where
\begin{equation}
\bm'=\left(\begin{array}{cc} m' \\
n'\\ \end{array}\right)=\left(\begin{array}{cc} dm -bn \\
-cm+an \\ \end{array}\right) \ ,
\end{equation}
and where $\Phi_{NS},G_{MN}^{NS}$ and $B_{MN}^{NS}$ correspond to
the $AdS_3\times S^3$ with NSNS flux.

 Now the equations of motion that follow from the action
have the form
\begin{eqnarray}\label{eqxM2}
&+&\partial_\alpha\left[G^{NS}_{MN}\partial_\beta x^N
g^{\beta\alpha} \sqrt{-\det g_{\alpha\beta}}\sqrt{m'^2+n'^2
e^{-2\Phi_{NS}}}\right]
-\nonumber \\
&-&\frac{1}{2}\partial_M G^{NS}_{KL}\partial_\alpha x^K
\partial_\beta
x^L \sqrt{m'^2+n'^2 e^{-2\Phi_{NS}}}+\nonumber \\
&+&m' H^{NS}_{MKN}\partial_\tau x^K\partial_\sigma x^N=0 \ ,
\nonumber \\
\end{eqnarray}
where
\begin{eqnarray}
H^{NS}_{MNK}&=&\partial_M B^{NS}_{NK}+\partial_N B^{NS}_{KM}+
\partial_K B^{NS}_{MN} \ . \nonumber \\
\end{eqnarray}
To proceed further we use the fact  that $AdS_3\times S^3$  is isomorphic to the group manifold
$G=SU(1,1)\times SU(2)$. Explicitly, let $g$ is the group element
from $G$. Then it is possible to write the metric (\ref{lineAdSNS})
as
\begin{equation}
G^{NS}_{MN}=L^2E_M^{ \ A}E_N^{ \ B}K_{AB} \ ,
\end{equation}
where for the group element $g\in G$ we have
\begin{equation}\label{defEM}
J\equiv g^{-1}dg=E_M^{ \  A}T_A dx^M \ ,
\end{equation}
where $T_A$ is the basis of Lie Algebra $\mG$ of the group $G$. Note
that $K_{AB}=\tr (T_A T_B)$. Further, from the definition
(\ref{defEM}) we obtain
\begin{equation}
dJ+J\wedge J=0
\end{equation}
that implies an important relation
\begin{equation}\label{relE}
\partial_M E_N^{ \ A}-\partial_N E_M^{ \ A}+f^A_{ \ BC}E_M^{  \
B}E_N^{  \ C}=0 \ ,
\end{equation}
where
\begin{equation}
[T_B,T_C]=T_A f^A_{ \ BC} \ .
\end{equation}
In case of the flux (\ref{HNSAdS}) we have following relation:
\begin{equation}\label{HFdef}
H^{NS}_{MNK}E^{M}_{ \ A}E^N_{ \ B} E^K_{ \ C}=L^2 f_{ABC} \  .
\end{equation}
With the help of (\ref{HFdef})  we can write
\begin{eqnarray}\label{EHx}
E^M_{ \ C}H_{MKL}\partial_\tau x^K \partial_\sigma x^L
&=& L^2 f_{CAB}J^A_\tau J^B_\sigma  \ , \nonumber \\
\end{eqnarray}
where  $E^M_{ \ A}$ is inverse to $E_M^{\ B}$ defined as
\begin{equation}
E^M_{ \ A}E_M^{ \ B}=\delta_A^B \ , \quad E^M_{  \ A}E_N^{ \ A}=
\delta^M_N \ .
\end{equation}
 Now with the help of
(\ref{relE}) and (\ref{EHx}) we can rewrite the equations of motion
(\ref{eqxM2}) to the form that contains the current
$J_\alpha^A=E_M^{ \ A}\partial_\alpha x^M$
\begin{eqnarray}\label{eqfinal}
 &&L^2T_{D1}K_{AB}\partial_\alpha[J_\beta^B g^{\beta\alpha}
\sqrt{-\det g_{\alpha\beta}} \sqrt{m'^2+n'^2 e^{-2\Phi_{NS}}}]+\nonumber \\
 &+&L^2T_{D1} m'  f_{ABC}J^B_\tau J^C_\sigma=0 \  , \quad g_{\alpha\beta}=
 K_{AB}J^A_\alpha J^B_\beta  \nonumber \\
\end{eqnarray}
that can be rewritten into the form
\begin{eqnarray}\label{conhjA}
\partial_\alpha \hJ^{A\alpha}=0 \ , \quad 
\hJ^{A\alpha}= L^2T_{D1}\left(J_\beta^A g^{\beta\alpha} \sqrt{-\det
g_{\alpha\beta}} \sqrt{m'^2+n'^2
e^{-2\Phi_{NS}}}+m'\epsilon^{\alpha\beta}J_\beta^A\right) \ .  \nonumber \\
\end{eqnarray}
 We see that the current $\hJ^{A\alpha}$ is conserved. Following
\cite{Kluson:2015lia}  we introduce an auxiliary metric
$\gamma_{\alpha\beta}$ that obeys the equation
\begin{equation}\label{Tab}
T_{\alpha\beta}\equiv\frac{1}{2}\gamma_{\alpha\beta}
\gamma^{\mu\nu}g_{\mu\nu}-g_{\alpha\beta}=0 \ .
\end{equation}
It is easy to see that this equation has solution
$\gamma_{\alpha\beta}=g_{\alpha\beta}$. If we further introduce
light-cone coordinates
\begin{equation}
\sigma^+=\frac{1}{2}(\tau+\sigma) \ , \quad
\sigma^-=\frac{1}{2}(\tau-\sigma) \
\end{equation}
we can rewrite the equation  (\ref{conhjA}) into the  form
\begin{equation}\label{curconser+-}
\partial_{+}\hJ^{A+}+\partial_{-}\hJ^{A-}=0 \ , \quad
\partial_\pm=\frac{\partial}{\partial \sigma^\pm} \ ,
\end{equation}
where
\begin{eqnarray}
\hJ^{A+}&=&\frac{1}{2}(\hJ^{A\tau}+\hJ^{A\sigma})= \nonumber
\\
&=&\frac{\sqrt{\lambda}}{2}\left[\sqrt{m'^2+e^{-2\Phi_{NS}}n'^
2}\sqrt{-\gamma}
\left(\gamma^{\tau\alpha}J_\alpha^A+\gamma^{\sigma\alpha}J_\alpha^A\right)+
m'(J_\sigma^A-J_\tau^A)\right]
 \ ,
\nonumber \\
\hJ^{A-}&=&\frac{1}{2}(\hJ^{A\tau}-\hJ^{A\sigma})= \nonumber \\
&=&\frac{\sqrt{\lambda}}{2}\left[\sqrt{m'^2+e^{-2\Phi_{NS}}n'^
2}\sqrt{-\gamma}(\gamma^{\tau
\alpha}J^A_\alpha-\gamma^{\sigma\alpha}J_\alpha^A)+
m'(J_\sigma^A+J_\tau^A)\right] \ , \nonumber \\
\end{eqnarray}
where $\sqrt{\lambda}=\frac{L^2}{2\pi\alpha'}$.  As the next step we
fix an auxiliary metric to have the form
$\gamma_{\alpha\beta}=\eta_{\alpha\beta} \ ,
\eta_{\alpha\beta}=\mathrm{diag}(-1,1)$ keeping in mind that
currents still have to obey the equation (\ref{Tab}). In this gauge
$\hJ^A_\pm$ simplify considerably and we obtain
\begin{eqnarray}\label{hjAgen}
\hJ^{A+}&=& -\frac{1}{2}\hJ^A_-=
\frac{\sqrt{\lambda}}{2}\left[J_\sigma^A\left(\sqrt{m'^2+n'^ 2
e^{-2\Phi_{NS}}}+m'\right)\right. \nonumber \\
&-& \left. J^A_\tau
\left(\sqrt{m'^2+e^{-2\Phi_{NS}}n'^2}+m'\right)\right] \ ,  \nonumber \\
\hJ^{A-}&=&-\frac{1}{2}\hJ^A_+=
-\frac{T_{D1}}{2}\left[J^A_\tau\left(\sqrt{m'^2+e^{-2\Phi_{NS}}n'^
2}
-m'\right) \right.   \nonumber \\
&+& \left. J_\sigma^A\left(\sqrt{m'^2+n'^ 2
e^{-2\Phi_{NS}}}-m'\right)\right]
\ , \nonumber \\
\end{eqnarray}
where we introduced the light-cone metric with
$\eta_{+-}=\eta_{-+}=-2 \ , \eta^{+-}=\eta^{-+}=-\frac{1}{2}$ so
that $\hJ^{A+}=\eta^{+-}\hJ^A_-=-\frac{1}{2}\hJ_-^A \ ,
\hJ^{A-}=\eta^{-+}\hJ^A_-=-\frac{1}{2}\hJ_+^A$. We see that for
\begin{equation}\label{omegaPicon}
 n'=0
\end{equation}
 the current $\hJ^A_+$ vanishes identically and the
equation (\ref{curconser+-}) gives
\begin{equation}\label{con+-}
\partial_{+}\hJ^A_-=0 \ , \quad  \hJ^A_-=2\sqrt{\lambda}m'
(J^A_\tau-J^A_\sigma) \ .
\end{equation}
Note that we can write $\hJ_-=\hJ^A_- T_A=2g^{-1}\partial_-g$. Then
from (\ref{con+-}) we obtain
\begin{equation}
\frac{1}{2}\partial_+ \hJ_-=-g^{-1}\partial_+g g^{-1}\partial_-g+
g^{-1}\partial_-\partial_+g
g^{-1}g=g^{-1}\partial_-[\partial_+gg^{-1}]g=0
\end{equation}
so that there is second current $\hJ_+=\partial_+gg^{-1}$ that obeys
the equation
\begin{equation}\label{con-+}
\partial_-\hJ_+=0 \ .
\end{equation}
Our result shows that $(a,c)-$string in the $(d,-b)-$mixed flux
background has two holomorphic and anti-holomorphic currents and
can be analyzed in the same way as WZW model \cite{Witten:1983ar}
using powerful conformal field theory techniques. On the other hand
 it is important to stress that $(m,n)-$string  in
 $(d,-b)-$background is still classically
  integrable  for any values of $m,n$ \cite{Kluson:2015lia}.

\section{Conclusion}\label{seventh}
Let us outline results derived in this paper. We found type IIB
supergravity solutions corresponding to $(p,q)-$string and
$(p,q)-$five brane backgrounds using $SL(2,Z)$ covariance of type
IIB superstring theory. We showed that these solutions have the
correct values of charges and also that the probe $(m,n)-$string in
this background can be mapped to the $(m',n')-$string in the
original fundamental or NS5-brane background where $m',n'$ are
integers whose values are predicted by $SL(2,Z)$ duality of type IIB
superstring theory. We also derived background that arises by
$SL(2,Z)$ transformation of the bound state of $Q_5$ NS5-branes and
$Q_1$ fundamental strings. Then we considered its near horizon limit
and argued that it leads to the $AdS_3\times S^3$ background with mixed
three form fluxes. Then we analyzed $(m,n)-$string  in given
background and we argued that for $m=a,n=c$ the string equations of
motion are equivalent to the conservation of two holomorphic and
antiholomorphic currents and hence this particular case can be
analyzed using  two dimensional conformal field theory. We
mean that this is a very interesting result that shows that
$(a,c)-$string is natural probe of $AdS_3\times S^3$ background with
$(d,-b)-$fluxes.

The extension of this work is as follows. It would be nice to
analyze solutions of $(m,n)-$string equations of motion in 
$AdS_3\times S^3$ background with mixed fluxes.
background. It would be also nice to analyze  integrability of
general $(m,n)-$string in this background
  in more details using the manifest covariant form of
$(m,n)-$string action. We hope to return to these problems in
future.

\vskip .5in \noindent {\bf Acknowledgement:}
\\
This work  was supported by the Grant Agency of the Czech Republic
under the grant P201/12/G028.

\end{document}